%% file: paper.tex
\newif\ifpreprint
\preprinttrue
\newif\ifcomment
\newif\ifnprl
\newif\ifdraft
\ifpreprint
\documentclass[ALICE,manyauthors]{cernphprep}
\usepackage[comma,square,numbers,sort&compress]{natbib}
\else
\documentclass[10pt,aps,prl,twocolumn,superscriptaddress,altaffilletter,nobibnotes,nofootinbib]{revtex4-1}
\usepackage{graphicx}
\fi
\usepackage[utf8]{inputenc}
\usepackage{hyperref}
\usepackage{lineno}
\usepackage{color}
\definecolor{dgreen}{cmyk}{1.,0.,1.,0.1}      
\definecolor{orange}{cmyk}{0.,0.353,1.,0.}    
\newcommand{\Figure}[1] {Figure~\ref{#1}}

\newcommand{\com}[1]    {\relax}

\def\dvers {v8}

\def\PbPb  {\mbox{Pb--Pb}}
\def\XeXe  {\mbox{Xe--Xe}}
\def\pPb   {\mbox{p--Pb}}

\def\pp    {\mbox{pp}}
\def\snn   {\mbox{$\sqrt{s_{_{\rm NN}}}$}}
\def\pt    {p_{\rm{T}}}
\def\scft  {{\rm SC}(4,2)_{\rm{3-sub}}}
\def\sctt  {\rm{SC}(3,2)_{\rm{3-sub}}}

\graphicspath{{.img/}}
\ifdraft
\linenumbers
\fi
\begin{document}%
\ifpreprint
\PHyear{2019}
\PHnumber{033}      
\PHdate{22 February}  
\begin{titlepage}
\fi
\title{Investigations of anisotropic flow using multi-particle\\ azimuthal correlations in pp, p--Pb,\\ Xe--Xe, and Pb--Pb collisions at the LHC}
\ifpreprint
\ShortTitle{Multi-particle correlations in pp, p--Pb, Xe--Xe and Pb--Pb}   
\Collaboration{ALICE Collaboration\thanks{See Appendix~\ref{app:collab} for the list of collaboration members}}
\ShortAuthor{ALICE Collaboration} 
\fi
\ifdraft
\begin{center}
\today\\ \color{red}DRAFT \dvers\ \hspace{0.3cm} \$Revision: 5573$\color{white}:$\$\color{black}\vspace{0.3cm}
\end{center}
\fi
\begin{abstract}
Measurements of anisotropic flow coefficients ($v_n$) and their cross-correlations using two- and multi-particle cumulant methods are reported  in collisions of pp at $\sqrt{s} = 13$ TeV, p--Pb at $\sqrt{s_{_{\rm NN}}} = 5.02$ TeV, Xe--Xe at $\sqrt{s_{_{\rm NN}}} = 5.44$ TeV, and Pb--Pb at $\sqrt{s_{_{\rm NN}}} = 5.02$ TeV recorded with the ALICE detector.
The multiplicity dependence of $v_n$ is studied in a very wide range from 20 to 3000 particles produced in the mid-rapidity region $|\eta|<0.8$ for the transverse momentum range $0.2 < p_{\rm T} < 3.0$ GeV/$c$. 
An ordering of the coefficients $v_2 > v_3 > v_4$ is found in pp and \pPb\ collisions, similar to that seen in large collision systems, while a weak $v_2$ multiplicity dependence is observed relative to nucleus--nucleus collisions in the same multiplicity range.
Using a novel subevent method, $v_{2}$ measured with four-particle cumulants is found to be compatible with that from six-particle cumulants in pp and \pPb\ collisions. 
The magnitude of the correlation between $v_n^2$ and $v_m^2$, evaluated with the symmetric cumulants SC$(m,n)$ is observed to be positive at all multiplicities for $v_2$ and $v_4$, while for $v_2$ and $v_3$ it is negative and changes sign for multiplicities below 100, which may indicate a different $v_n$ fluctuation pattern in this multiplicity range.
The observed long-range multi-particle azimuthal correlations in high multiplicity pp and \pPb\ collisions can neither be described by PYTHIA 8 nor by IP-Glasma+MUSIC+UrQMD model calculations, and hence provide new insights into the understanding of collective effects in small collision systems.
\end{abstract}
\ifpreprint
\end{titlepage}
\setcounter{page}{2}
\else
\maketitle
\fi
\ifnprl
\textbf{[Introduction]}
\fi
Experiments investigating ultra-relativistic collisions of heavy ions intend to explore a deconfined state of quarks and gluons, the Quark--Gluon Plasma (QGP). Azimuthal correlations of final state particles over a wide range in pseudorapidity relative to the collision symmetry plane $\Psi_n$~($n\ge1$), whose magnitudes are quantified by flow coefficients $v_n$, provide important information into the matter created in these collisions~\cite{Ollitrault:1992bk, Voloshin:2008dg, Heinz:2013th}. 
Extensive measurements of $v_n$ for inclusive~\cite{Adam:2016ows, Adam:2016izf, Sirunyan:2017fts, Acharya:2018lmh, Acharya:2018ihu, Aaboud:2018ves} and identified hadrons~\cite{Acharya:2018zuq} were performed for \XeXe\ and \PbPb\ collisions at the Large Hadron Collider~(LHC). 
These studies, together with quantitative descriptions by hydrodynamic calculations, have enabled an extraction of the properties of the QGP~\cite{Song:2017wtw}, revealing that it behaves as a nearly-perfect fluid with a shear viscosity over entropy density ratio $\eta/s$ close to the universal lower limit 1/(4$\pi$) from Ads/CFT~\cite{Kovtun:2004de}. 
Recently, significant progress has also been achieved by measuring correlations between different flow coefficients and symmetry planes~\cite{Adam:2015eta, ALICE:2016kpq, Acharya:2017zfg, Acharya:2017ino, Acharya:2017gsw, Aaboud:2017tql, Sirunyan:2017fts,Acharya:2018lmh}.
In particular, the correlation strength between different flow coefficients~$v_m^2$ and $v_n^2$, quantified by symmetric cumulants $SC(m,n)$~\cite{Bilandzic:2013kga},was found to be sensitive to the temperature dependence of $\eta/s$ and the initial conditions~\cite{ALICE:2016kpq}. The experimental measurements of $SC(m,n)$, together with $v_n$, thus provide tighter constraints on theoretical models than the individual flow coefficients alone~\cite{ALICE:2016kpq,Acharya:2017gsw}.

Striking similarities between numerous observables, thought to indicate the emergence of a QGP, were observed across different collision systems at both RHIC and LHC energies, when compared at similar multiplicity of produced particles within a specific phase space~\cite{Loizides:2016tew,PHENIX:2018lia,Adam:2019woz}.
The ``ridge'' structure measured using two-particle correlations as a function of the pseudorapidity difference $\Delta\eta$ and the azimuthal angle difference $\Delta\varphi$, which in heavy-ion collisions results from anisotropic flow, was also observed in high multiplicity p--A and pp collisions~\cite{Aaboud:2016yar}. 
In addition, measurements of azimuthal correlations using multi-particle cumulants revealed signatures collective effects in small systems, such as a negative four-particle cumulant $c_2\{4\}$~\cite{Abelev:2014mda,Khachatryan:2015waa, Khachatryan:2016txc, Aaboud:2017acw, Aaboud:2017blb}.

Whether the observed similarities between small (\pp\ and p--A) and large (A--A) collision systems arise from the same physics mechanism is under intense debate.
Besides hydrodynamic descriptions~\cite{Bozek:2011if,Bzdak:2013zma,Mantysaari:2017cni,Weller:2017tsr,Zhao:2017rgg}, calculations from transport models~\cite{Bzdak:2014dia, Kurkela:2018ygx, Nie:2018xog}, hadronic rescattering~\cite{Zhou:2015iba,Romatschke:2015dha}, a string rope and shoving mechanism~\cite{Bierlich:2017vhg}, as well as initial stage effects~\cite{Welsh:2016siu, Dusling:2017dqg, Blok:2018xes}, have been investigated.

We report measurements of $v_n$ and $SC(m,n)$ as a function of produced particle multiplicity across small and large collision systems. 
These measurements provide information on the collective effects observed in all systems, which can be studied via long-range multi-particle correlations.
A significant extension of recent studies~\cite{Aaboud:2017blb,Sirunyan:2017uyl,Aaboud:2018syf} is achieved by adding new results of $v_2$ and SC$(m,n)$ for all available collision systems at the LHC, together with a comprehensive comparison to the available models ranging from non-flow dominated (PYTHIA 8) to the state of the art hydrodynamic model calculations.
They rely on a new technique of performing multi-particle long range correlations named the \textit{subevent~method}~\cite{Jia:2017hbm,Huo:2017nms}, which further minimises biases from few particle correlations such as resonances and jets, usually called non-flow, which are not associated with a collision symmetry plane. 

\ifnprl
\textbf{[Data sets]}
\fi
The analyzed data are from collisions of \pp\ at $\sqrt{s}=13$~TeV, \pPb\ at $\snn=$ 5.02 TeV, \XeXe\ at $\snn=5.44$~TeV and \PbPb\ at $\snn =$ 5.02 TeV.
They were recorded with the ALICE detector~\cite{Aamodt:2008zz,Abelev:2014ffa} during the years 2015, 2016 and 2017.
Minimum bias events were triggered using a coincidence signal in the two scintillator arrays of the V0 detector, V0A and V0C, which cover the pseudorapidity ranges $2.8<\eta<5.1$ and $-3.7<\eta<-1.7$, respectively~\cite{Abbas:2013taa}. 
A dedicated trigger was used in pp collisions to select high-multiplicity events based on the amplitude in both arrays of the V0 detector.
The trigger selected approximately 0.1\% of events with the largest multiplicity in the V0 acceptance. 
The corresponding average multiplicity is at least 4 times larger than in minimum bias collisions. 
In comparison to minimum-bias collisions, the selection of high-multiplicity events based on forward multiplicity suppresses the non-flow contribution to $v_n$ at mid-rapidity by suppressing jet correlations.

Only events with a reconstructed primary vertex $Z_{\rm vtx}$ within $\pm10$~cm from the nominal interaction point were selected.
A removal of background events from e.g.\ beam interaction with the residual gas molecules in the beam pipe and pileup events was performed based on the information from the Silicon Pixel Detector (SPD) and V0 detectors.
A sample of $310\times10^6$ high-multiplicity \pp, $230\times10^6$ minimum bias \pPb, $1.3\times10^6$ \XeXe\ and $55\times10^6$ \PbPb\ collisions that passed the event selection criteria was used for the analysis.
 
The charged tracks were reconstructed using the Inner Tracking System (ITS)~\cite{Aamodt:2010aa} and the Time Projection Chamber (TPC)~\cite{Alme:2010ke}. 
Only tracks with more than 70 clusters in the TPC (out of a maximum of 159) were selected. 
A selection requiring the pseudorapidity to be within $-0.8<\eta<0.8$ ensured a high track reconstruction efficiency of 80\%.
Tracks with a transverse momentum $\pt < 0.2$~GeV/$c$ and $\pt > 3.0$~GeV/$c$ were rejected due to low tracking efficiency and to reduce the contribution from jets, respectively. 
A criterion on the maximum distance of closest approach~(DCA) of the track to the collision point of less than $2$~cm in longitudinal direction and a $\pt$-dependent selection in the transverse direction, ranging from $0.2$~cm at $\pt=0.2$ GeV/$c$ down to $0.02$~cm at $\pt=3.0$ GeV/$c$, was applied leading to a residual contamination from secondaries between 1-3\%.

\ifnprl
\textbf{[Methods]}
\fi
The results were calculated from two- and multi-particle azimuthal correlations using the generic framework~\cite{Bilandzic:2013kga}, recently extended to include the subevent method~\cite{Huo:2017nms}. 
The ranges of the subevents were chosen to be $(-0.8, 0)$ and $(0, 0.8)$ for the 2-subevent, and $(-0.8, -0.4)$, $(-0.4, 0.4)$  and $(0.4, 0.8)$ for 3-subevent measurements. 

A correction dependent on $\eta$ and $Z_{\rm{vtx}}$ was applied to account for azimuthal non-uniformity. 
The correction for tracking inefficiencies was obtained from Monte Carlo simulations as a function of $\pt$, $\eta$ and $Z_{\rm{vtx}}$ from generated particles and from tracks reconstructed from a GEANT3 simulation~\cite{geant3}. 
The systematic uncertainties were estimated as follows.
The contribution from the event selection was examined by narrowing the selection on $Z_{\rm{vtx}}$ to $\pm 5$ cm.
The track reconstruction biases were evaluated by tightening the selection criteria on the DCA in both the longitudinal and transverse directions, by increasing the required minimum number of TPC clusters in the track reconstruction, and by comparing the results to those obtained with tracks having different requirements regarding the role of the ITS.
The uncertainty from the Monte Carlo closure test was estimated by comparing calculations at the event generator level with the simulation output after the full reconstruction.
The individual contributions were summed in quadrature to form the systematic uncertainties, ranging between 1-6\% for the two-particle cumulant, and 10-17\% for the multi-particle cumulant results. 
The results are reported as a function of the number of produced charged particles $N_{\rm ch} (|\eta| < 0.8, 0.2 < \pt < 3.0$ GeV/$c$). 

\begin{figure}[htbp!]
\begin{center}
\includegraphics[width=0.9\linewidth]{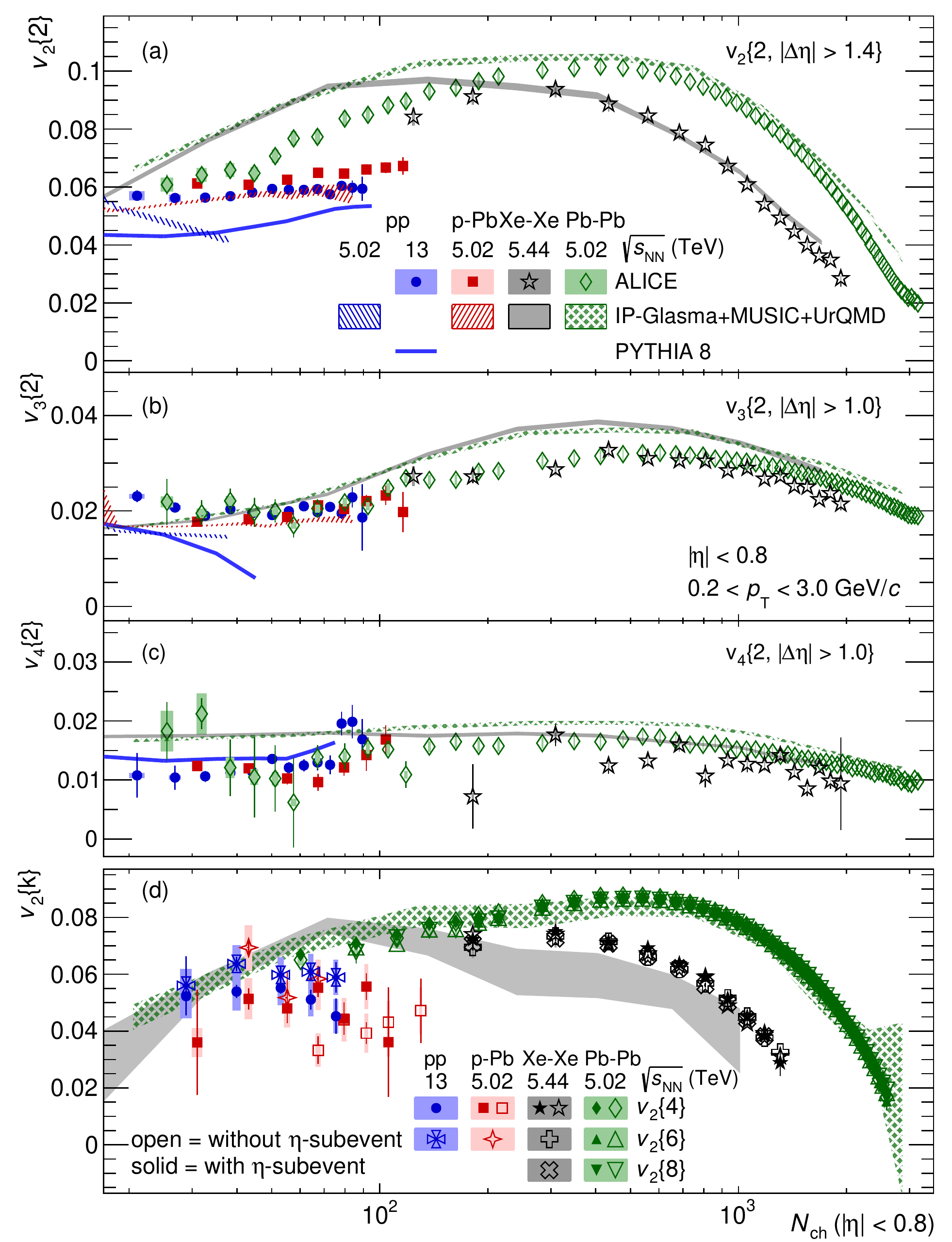}
\caption{(coloured online) Multiplicity dependence of $v_n\{k\}$ for \pp, \pPb, \XeXe\ and \PbPb\ collisions. 
	Statistical uncertainties are shown as vertical lines and systematic uncertainties as filled boxes.
	Data are compared with PYTHIA 8.210 Monash 2013~\cite{Sjostrand:2014zea} simulations (solid lines) of pp collisions at $\sqrt{s} = 13$ TeV and IP-Glasma+MUSIC+UrQMD~\cite{Schenke:2013dpa,Mantysaari:2017cni} calculations of pp, \pPb, \PbPb\ collisions at $\snn = 5.02$ TeV and \XeXe\ collisions at $\snn = 5.44$ TeV (filled bands). The width of the band represents the statistical uncertainty of the model.
         (a), (b) and (c): $v_2$, $v_3$ and $v_4$ measured using two-particle cumulants with a pseudorapidity separation $|\Delta\eta| > 1.4$, 1.0 and 1.0, respectively. 
         (d): $v_2$ measured using multi-particle cumulants, with the 3-subevent method for the four-particle cumulant, and 2-subevent method for higher order cumulants in \PbPb\ collisions.}
\label{fig1} 
\end{center}
\end{figure}

\ifnprl
\textbf{[Results]}
\fi
\Figure{fig1} presents the measurements of anisotropic flow coefficients $v_n\{k\}$ of order $n$, obtained from $k$-particle correlations, in \pp, \pPb, \XeXe\ and \PbPb\ collisions.
The collision energies are similar except for \pp\ collisions where no collision energy dependence of the integrated $v_n$ is expected~\cite{Aaboud:2017acw}.

Figures \ref{fig1}~(a), \ref{fig1}~(b) and \ref{fig1}~(c) show $v_2$, $v_3$ and $v_4$ measured using two-particle~($k=2$) cumulants with a pseudorapidity separation $|\Delta\eta| > 1.4$, 1.0 and 1.0 respectively, chosen to suppress non-flow contributions. 
Due to the limited statistics of the \pp\ data sample, the $|\Delta\eta|$ separation in case of $v_3$ and $v_4$ was reduced to 1.0, consistently across all collision systems.
A pronounced multiplicity dependence of $v_2$ is observed in the flow dominated collision systems (\PbPb\ and \XeXe) as a result of the medium response to the eccentricity of the initial overlap region of the colliding nuclei. 
The \PbPb\ data exhibits larger $v_2$ values than the \XeXe\ data, but they are compatible for $N_{\rm ch} < 200$.
An ordering of $v_2 > v_3 > v_4$ is observed in large systems except for the very high multiplicities, where $v_2 \approx v_3$. 
At low multiplicity, the magnitudes of $v_n$ are similar to those measured in \pp\ and \pPb\ collisions. 
The measurements from large systems are compared with calculations using IP-Glasma initial conditions, MUSIC hydrodynamic model, and the UrQMD model for hadronic rescatterings~\cite{Schenke:2013dpa,Mantysaari:2017cni}. 
The calculations qualitatively describe all the $v_n$ measurements except for $N_{\rm ch} < 200$ where they overestimate the $v_2$.

In small collision systems, all the $v_n$ coefficients exhibit a weak dependence on multiplicity.
The trend and magnitudes, particularly for $v_2$, cannot be explained solely by model calculations without collective effects.
This can be demonstrated by the comparison with predictions from PYTHIA 8~\cite{Sjostrand:2014zea}, computed with a similar multiplicity definition as the experimental results from pp collisions.
The  ordering of $v_n$ in pp collisions for all multiplicities is the same as in large collision systems ($v_2 > v_3 > v_4$) and is not described by PYTHIA~8 where $v_2 > v_4 > v_3$ for $N_{\rm ch} > 30$.
These observations suggest the presence of effects other than just non-flow correlations at multiplicities larger than about $2$--$3$ times the minimum bias value of $\langle N_{\rm ch}\rangle \approx 10$ in \pp\ and $\langle N_{\rm ch}\rangle \approx 24$ in \pPb\ collisions. 
In \pPb\ collisions these conclusions are further supported by the qualitative agreement with the IP-Glasma+MUSIC+UrQMD model calculations.
Nevertheless, the hydrodynamic model reveals a strong decrease of $v_2$ with multiplicity in \pp\ collisions, which is in stark contrast with the data. 
A further non-flow suppression with a larger $|\Delta\eta|$ separation in the experimental results of \pPb\ collisions, or improvements in the phenomenological description, might help to reach a quantitative agreement.

\Figure{fig1}~(d) shows measurements of $v_2\{k\}$ using cumulants with a number $k = 4, 6$ and 8 particles. 
Measurements of $v_2\{4\}$ with the 3-subevent method, and of $v_2\{6\}$ and $v_2\{8\}$ in \PbPb\ collisions with the 2-subevent method, are also presented.
Compared to $v_2\{2\}$, multi-particle cumulants are less influenced by non-flow effects, since the latter usually  involve only a few particles. 
No further non-flow suppression was observed by increasing the $|\Delta\eta|$ separation between the subevents in the multi-particle cumulant measurements.
In \XeXe\ and \PbPb\ collisions, characteristic patterns of long-range multi-particle correlations, such as consistent results from the standard and subevent methods ($v_2\{4\} \approx v_2\{4\}_{\rm{3-sub}}$, $v_2\{6\} \approx v_2\{6\}_{\rm{2-sub}}$ and $v_2\{8\} \approx v_2\{8\}_{\rm{2-sub}}$), and compatible measurements of $v_2$ with multi-particle cumulants ($v_2\{4\} \approx v_2\{6\} \approx v_2\{8\}$) are found, signalling a negligible contribution from non-flow correlations and the dominance of collective effects.
Moreover, a good agreement of $v_2\{4\}$ between data and calculations from IP-Glasma+MUSIC+UrQMD~\cite{Schenke:2013dpa,Mantysaari:2017cni} model is found for \PbPb\ collisions down to $N_{\rm ch} \approx 200$.
The same model prediction, which does not include any tuning of its parameters to other collision systems, underestimates the $v_2\{4\}$ from \XeXe\ collisions by about 15--20\%. 

In \pPb\ collisions, a further non-flow suppression with the 3-subevent method leads to a decrease of the cumulant $c_2\{4\} > c_2\{4\}_{\rm 3-sub}$, which due to the relation $v_2\{4\} = \sqrt[4]{-c_2\{4\}}$ corresponds up to a $2\sigma$ increase $v_2\{4\} < v_2\{4\}_{\rm 3-sub}$.
The 3-subevent method allows for a measurement of a real-valued $v_2\{4\}_{\rm 3-sub}$ at a lower $N_{\rm ch}$ than the standard $v_2\{4\}$ measurement, making it possible to study collectivity at even lower multiplicities.
Genuine multi-particle correlations in \pPb\ collisions are indicated by consistent results of $v_2\{4\}$ and $v_2\{6\}$. In pp collisions, significant non-flow contributions to the four-particle cumulant ($c_2\{4\} > 0$) prevent the extraction of a real-valued $v_2\{4\}$. 
However, a measurement of the real-valued $v_2\{4\}_{\rm{3-sub}}$ is possible with the 3-subevent method.
Similarly as for $v_2\{2, |\Delta\eta| > 1.4\}$, the $v_2\{4\}_{\rm{3-sub}}$ exhibits only a weak dependence on multiplicity.
These results confirm the existence of long-range multi-particle correlations in \pp\ and \pPb\ collisions at multiplicities $N_{\rm ch} \geq 30$. 
PYTHIA 8 calculations, which do not contain genuine long-range multi-particle correlations, do not give a real valued $v_{2}\{4\}$ even with the subevent method~\cite{Jia:2017hbm}.
The superSONIC~\cite{Weller:2017tsr} and iEBE-VISHNU~\cite{Zhao:2017rgg} hydrodynamic models, which can quantitatively describe all available two-particle correlation measurements in \pp, \pPb\ and \PbPb\ collisions, cannot reproduce the four-particle cumulants with the currently used initial state model, not even on a qualitative level.
Another model with initial-state calculations predicts the multi-particle cumulants with correct signs and a weak dependence on the saturation scale $Q_s^2$, but the predictions are 10 times larger than what is observed in the data and there is no direct connection of the $Q_s^2$ to the experimentally measured number of produced charged particles~\cite{Dusling:2017dqg}. 
Therefore, with $v_n$ measurements alone it is not completely clear whether the origin of the apparent collectivity observed in small collision systems is the same as in large collision systems.

\begin{figure}[htbp!]
\begin{center}
\includegraphics[width=0.8\linewidth]{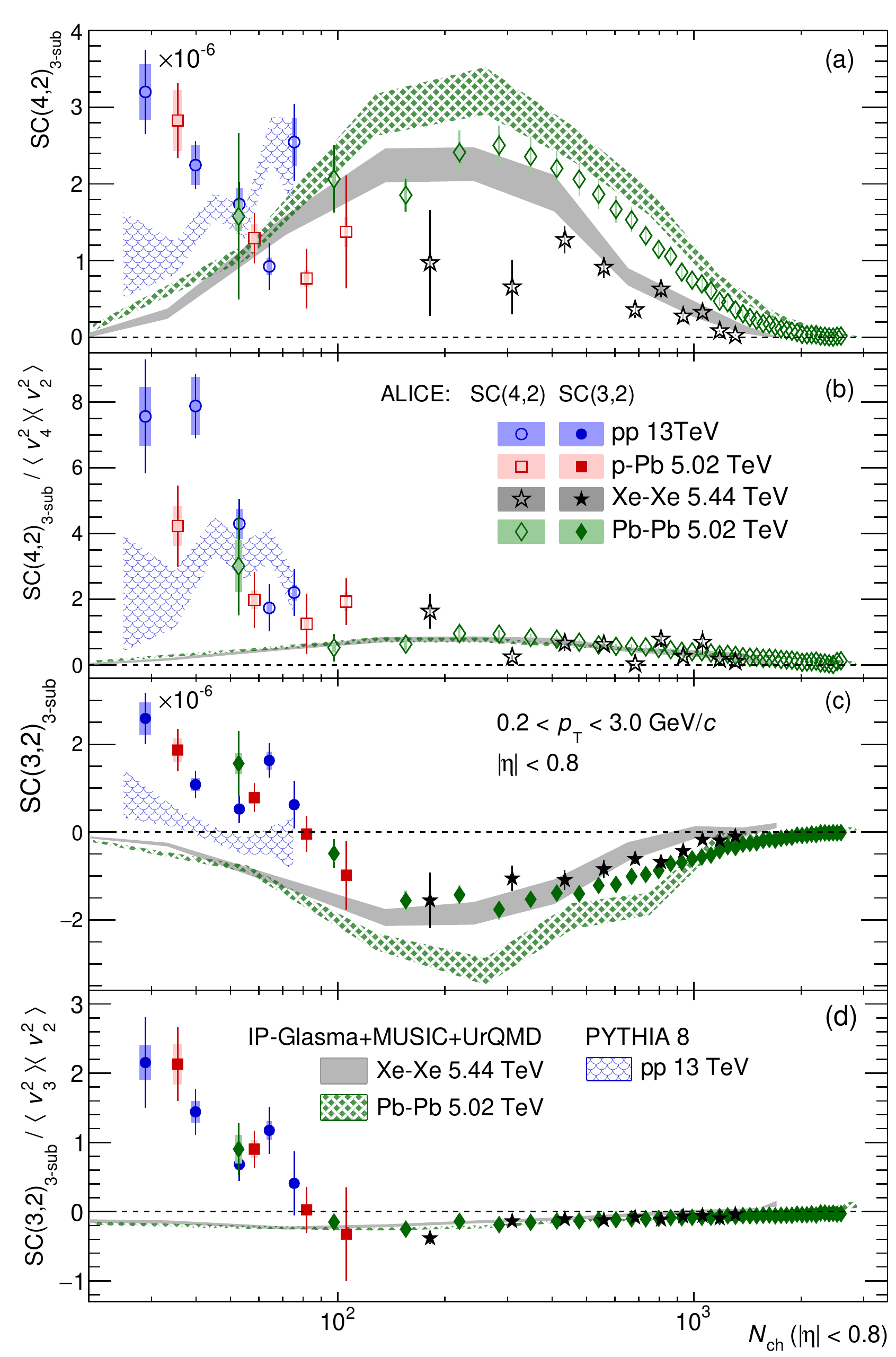}
\caption{(coloured online) Multiplicity dependence of the (a) and (c) symmetric cumulant $SC(m,n)_{\rm{3-sub}}$ and (b) and (d) normalized ratio SC$(m,n)_{\rm{3-sub}} / \langle v_m^2\rangle\langle v_n^2\rangle$ for \pp, \pPb, \XeXe\ and \PbPb\ collisions. Observables in the denominator are obtained from the $v_2\{2, |\Delta\eta| > 1.4\}$ and $v_n\{2, |\Delta\eta| > 1.0\}$ for higher harmonics. Statistical uncertainties are shown as vertical lines and systematic uncertainties as filled boxes. The measurements in large collision systems are compared with the IP-Glasma+MUSIC+UrQMD~\cite{Schenke:2013dpa,Mantysaari:2017cni} calculations and results in pp collisions are compared with the PYTHIA 8 model~\cite{Sjostrand:2014zea}.}
\label{fig2}
\end{center}
\end{figure}

Further information about the origin of the observed collectivity can be obtained from symmetric cumulants SC$(m,n)$, which quantify the correlation between $v_m^2$ and $v_n^2$.
Figures~\ref{fig2}~(a) and \ref{fig2}~(c) present the multiplicity dependence of SC$(m,n)$ measured with the 3-subevent method. 
In Fig.~\ref{fig2}~(a), a positive $\scft$ is observed in large systems over the entire multiplicity range, similar to what was measured previously in \PbPb\ collisions at 2.76~TeV~\cite{ALICE:2016kpq,Acharya:2017gsw} without the subevent method. 
The trend is reproduced by the IP-Glasma+MUSIC+UrQMD~\cite{Schenke:2013dpa,Mantysaari:2017cni} calculations.
A similar positive $\scft$ is observed both in \pp\ and \pPb\ collisions, as was also found in~\cite{Aaboud:2018syf}. 
The measurements in pp collisions are compared with PYTHIA 8~\cite{Sjostrand:2014zea}, which shows a decrease of $\scft$ with decreasing multiplicity, different from what is seen in data. 
Calculations~\cite{He:2015hfa,Dusling:2017dqg} with initial state correlations or parton-escape mechanism can qualitatively or even semi-quantitatively describe the \pPb\ data. 
We note that the results from the initial state model~\cite{Dusling:2017dqg} were calculated as a function of variables that cannot be directly computed from experimental data.

An anti-correlation between $v_2^2$ and $v_3^2$ is implied by the negative $\sctt$ observed in \XeXe\ and \PbPb\ collisions for $N_{\rm ch} > 100$ in Fig.~\ref{fig2} (c), similarly as in~\cite{ALICE:2016kpq,Acharya:2017gsw}.
There is a hint of a change to a positive sign of $\sctt$ in \PbPb\ collisions below multiplicities of $N_{\rm ch} \approx 100$. 
This tendency is observed at even lower multiplicities in small collision systems, suggesting a common positive correlation between $v_2^2$ and $v_3^2$ among collision systems of different sizes.
Such a behaviour is not observed for small collision systems with a larger $\eta$ acceptance~\cite{Aaboud:2018syf}, where $\sctt$ remains negative in the whole multiplicity range.
One possible explanation is the different contributions from non-flow effects.
The IP-Glasma+MUSIC+UrQMD~\cite{Schenke:2013dpa,Mantysaari:2017cni} calculations for \XeXe\ and \PbPb\ collisions reproduce the negative correlation at large multiplicities. 
This negative sign persists in simulations down to the lowest multiplicities.
PYTHIA 8~\cite{Sjostrand:2014zea} fails to quantitatively describe the results from pp collisions, but it does qualitatively reproduce the trend of the data.
No hydrodynamic calculations of SC$(m,n)$ in small systems are currently available. 
Nevertheless, calculations based on initial state correlations in~\cite{Welsh:2016siu,Dusling:2017dqg} reflect the crossing from negative to positive SC$(3,2)$ in \pPb\ collisions, whereas a positive correlation is predicted in \pp\ collisions~\cite{Welsh:2016siu}.

While SC$(m,n)$ encodes information on both the magnitude of and correlation between the flow coefficients, in the absence of non-flow the latter can be accessed directly by dividing SC$(m,n)_{\rm{3-sub}}$ by the corresponding flow coefficients $\langle v_m^2\rangle\langle v_n^2\rangle$. 
The normalised ratios, shown in Figs.~\ref{fig2}~(b) and \ref{fig2}~(d), indicate that the correlation between flow coefficients is possibly the same between different collision systems at the same $N_{\rm ch}$, and reveals a large increase in magnitude in the correlation strength for collisions with $N_{\rm ch} < 100$ compared to higher multiplicities. 
While this may be indicative of a different fluctuation pattern at low multiplicity, non-flow effects likely persist in this region based on the observed finite values of PYTHIA 8 calculations.
Such effects make the interpretation of an increase of the normalised ratio significantly less straightforward and requires further study.

\ifnprl
\textbf{[Summary]}
\fi
In summary, we have presented the measurements of flow coefficients $v_n\{k\}$ and symmetric cumulants SC$(m,n)$ as a function of the produced particle multiplicity in small (\pp, \pPb) and large (\XeXe, \PbPb) collision systems.
In \pp\ and \pPb\ collisions, an ordering $v_{2} > v_{3} > v_{4}$ and a weak dependence of $v_n$ on the multiplicity, is observed.
The values of $v_n$ from \pp\ and \pPb\ collisions are compatible with heavy-ion collisions at low multiplicities. 
These first ALICE measurements of $v_2$ using multi-particle cumulants in small collision systems are found to be compatible with each other after a suppression of non-flow contributions with the subevent method.
Positive values of $\scft$ are seen in all four collision systems~(\pp, \pPb, \XeXe\ and \PbPb).
The observed anti-correlation between $v_2^2$ and $v_3^2$ measured with $\sctt$ in large collision systems seems to evolve into a positive correlation at low multiplicity.
A similar sign change is indicated also in \pp\ and \pPb\ collisions.
Thus, the different systems exhibit a similar SC$(m,n)$ at the same $N_{\rm ch}$, and below $N_{\rm ch} < 100$ reveal a large variation of the correlation strength and/or an increasing contribution of non-flow.
The measurements in pp collisions can not be reproduced by the PYTHIA 8 model.
The hydrodynamic description with the IP-Glasma+MUSIC+UrQMD calculations shows rather good agreement with data in \PbPb, \XeXe\ and \pPb\ collisions, but fails to describe the measurements in pp collisions, where applicable. 
The presented data provide new information about origin of the observed collectivity, and provides key constraints to the various approaches for modeling collectivity in small systems.

\newenvironment{acknowledgement}{\relax}{\relax}
\begin{acknowledgement}
\section*{Acknowledgements}
\ifdraft
\else
\input{fa_2019-02-07.tex}
\fi
\end{acknowledgement}
\bibliographystyle{utphys}
\bibliography{biblio}

\newpage
\section{The ALICE Collaboration}
\label{app:collab}
\ifdraft
\else
\input{2019-02-07-Alice_Authorlist_2019-Feb-07.tex}
\fi

\end{document}

%% file: fa_2019-02-07.tex

The ALICE Collaboration would like to thank all its engineers and technicians for their invaluable contributions to the construction of the experiment and the CERN accelerator teams for the outstanding performance of the LHC complex.
The ALICE Collaboration gratefully acknowledges the resources and support provided by all Grid centres and the Worldwide LHC Computing Grid (WLCG) collaboration.
The ALICE Collaboration acknowledges the following funding agencies for their support in building and running the ALICE detector:
A. I. Alikhanyan National Science Laboratory (Yerevan Physics Institute) Foundation (ANSL), State Committee of Science and World Federation of Scientists (WFS), Armenia;
Austrian Academy of Sciences, Austrian Science Fund (FWF): [M 2467-N36] and Nationalstiftung f\"{u}r Forschung, Technologie und Entwicklung, Austria;
Ministry of Communications and High Technologies, National Nuclear Research Center, Azerbaijan;
Conselho Nacional de Desenvolvimento Cient\'{\i}fico e Tecnol\'{o}gico (CNPq), Universidade Federal do Rio Grande do Sul (UFRGS), Financiadora de Estudos e Projetos (Finep) and Funda\c{c}\~{a}o de Amparo \`{a} Pesquisa do Estado de S\~{a}o Paulo (FAPESP), Brazil;
Ministry of Science \& Technology of China (MSTC), National Natural Science Foundation of China (NSFC) and Ministry of Education of China (MOEC) , China;
Croatian Science Foundation and Ministry of Science and Education, Croatia;
Centro de Aplicaciones Tecnol\'{o}gicas y Desarrollo Nuclear (CEADEN), Cubaenerg\'{\i}a, Cuba;
Ministry of Education, Youth and Sports of the Czech Republic, Czech Republic;
The Danish Council for Independent Research | Natural Sciences, the Carlsberg Foundation and Danish National Research Foundation (DNRF), Denmark;
Helsinki Institute of Physics (HIP), Finland;
Commissariat \`{a} l'Energie Atomique (CEA), Institut National de Physique Nucl\'{e}aire et de Physique des Particules (IN2P3) and Centre National de la Recherche Scientifique (CNRS) and Rl\'{e}gion des  Pays de la Loire, France;
Bundesministerium f\"{u}r Bildung, Wissenschaft, Forschung und Technologie (BMBF) and GSI Helmholtzzentrum f\"{u}r Schwerionenforschung GmbH, Germany;
General Secretariat for Research and Technology, Ministry of Education, Research and Religions, Greece;
National Research, Development and Innovation Office, Hungary;
Department of Atomic Energy Government of India (DAE), Department of Science and Technology, Government of India (DST), University Grants Commission, Government of India (UGC) and Council of Scientific and Industrial Research (CSIR), India;
Indonesian Institute of Science, Indonesia;
Centro Fermi - Museo Storico della Fisica e Centro Studi e Ricerche Enrico Fermi and Istituto Nazionale di Fisica Nucleare (INFN), Italy;
Institute for Innovative Science and Technology , Nagasaki Institute of Applied Science (IIST), Japan Society for the Promotion of Science (JSPS) KAKENHI and Japanese Ministry of Education, Culture, Sports, Science and Technology (MEXT), Japan;
Consejo Nacional de Ciencia (CONACYT) y Tecnolog\'{i}a, through Fondo de Cooperaci\'{o}n Internacional en Ciencia y Tecnolog\'{i}a (FONCICYT) and Direcci\'{o}n General de Asuntos del Personal Academico (DGAPA), Mexico;
Nederlandse Organisatie voor Wetenschappelijk Onderzoek (NWO), Netherlands;
The Research Council of Norway, Norway;
Commission on Science and Technology for Sustainable Development in the South (COMSATS), Pakistan;
Pontificia Universidad Cat\'{o}lica del Per\'{u}, Peru;
Ministry of Science and Higher Education and National Science Centre, Poland;
Korea Institute of Science and Technology Information and National Research Foundation of Korea (NRF), Republic of Korea;
Ministry of Education and Scientific Research, Institute of Atomic Physics and Ministry of Research and Innovation and Institute of Atomic Physics, Romania;
Joint Institute for Nuclear Research (JINR), Ministry of Education and Science of the Russian Federation, National Research Centre Kurchatov Institute, Russian Science Foundation and Russian Foundation for Basic Research, Russia;
Ministry of Education, Science, Research and Sport of the Slovak Republic, Slovakia;
National Research Foundation of South Africa, South Africa;
Swedish Research Council (VR) and Knut \& Alice Wallenberg Foundation (KAW), Sweden;
European Organization for Nuclear Research, Switzerland;
National Science and Technology Development Agency (NSDTA), Suranaree University of Technology (SUT) and Office of the Higher Education Commission under NRU project of Thailand, Thailand;
Turkish Atomic Energy Agency (TAEK), Turkey;
National Academy of  Sciences of Ukraine, Ukraine;
Science and Technology Facilities Council (STFC), United Kingdom;
National Science Foundation of the United States of America (NSF) and United States Department of Energy, Office of Nuclear Physics (DOE NP), United States of America.

%% file: 2019-02-07-Alice_Authorlist_2019-Feb-07.tex

\begingroup
\small
\begin{flushleft}
S.~Acharya\Irefn{org141}\And 
D.~Adamov\'{a}\Irefn{org93}\And 
S.P.~Adhya\Irefn{org141}\And 
A.~Adler\Irefn{org74}\And 
J.~Adolfsson\Irefn{org80}\And 
M.M.~Aggarwal\Irefn{org98}\And 
G.~Aglieri Rinella\Irefn{org34}\And 
M.~Agnello\Irefn{org31}\And 
N.~Agrawal\Irefn{org10}\And 
Z.~Ahammed\Irefn{org141}
% ...
\And
\renewcommand\labelenumi{\textsuperscript{\theenumi}~}

\section*{Affiliation notes}
\renewcommand\theenumi{\roman{enumi}}
\begin{Authlist}
\item \Adef{org*}Deceased
\item \Adef{orgI}Dipartimento DET del Politecnico di Torino, Turin, Italy
\item \Adef{orgII}M.V. Lomonosov Moscow State University, D.V. Skobeltsyn Institute of Nuclear, Physics, Moscow, Russia
\item \Adef{orgIII}Department of Applied Physics, Aligarh Muslim University, Aligarh, India
\item \Adef{orgIV}Institute of Theoretical Physics, University of Wroclaw, Poland
\end{Authlist}

\section*{Collaboration Institutes}
\renewcommand\theenumi{\arabic{enumi}~}
\begin{Authlist}
\item \Idef{org1}A.I. Alikhanyan National Science Laboratory (Yerevan Physics Institute) Foundation, Yerevan, Armenia
\item \Idef{org2}Bogolyubov Institute for Theoretical Physics, National Academy of Sciences of Ukraine, Kiev, Ukraine
\item \Idef{org3}Bose Institute, Department of Physics  and Centre for Astroparticle Physics and Space Science (CAPSS), Kolkata, India
\item \Idef{org4}Budker Institute for Nuclear Physics, Novosibirsk, Russia
\item \Idef{org5}California Polytechnic State University, San Luis Obispo, California, United States
\item \Idef{org6}Central China Normal University, Wuhan, China
\item \Idef{org7}Centre de Calcul de l'IN2P3, Villeurbanne, Lyon, France
\item \Idef{org8}Centro de Aplicaciones Tecnol\'{o}gicas y Desarrollo Nuclear (CEADEN), Havana, Cuba
\item \Idef{org9}Centro de Investigaci\'{o}n y de Estudios Avanzados (CINVESTAV), Mexico City and M\'{e}rida, Mexico
\item \Idef{org10}Centro Fermi - Museo Storico della Fisica e Centro Studi e Ricerche ``Enrico Fermi', Rome, Italy
\item \Idef{org11}Chicago State University, Chicago, Illinois, United States
\item \Idef{org12}China Institute of Atomic Energy, Beijing, China
\item \Idef{org13}Chonbuk National University, Jeonju, Republic of Korea
\item \Idef{org14}Comenius University Bratislava, Faculty of Mathematics, Physics and Informatics, Bratislava, Slovakia
\item \Idef{org15}COMSATS University Islamabad, Islamabad, Pakistan
\item \Idef{org16}Creighton University, Omaha, Nebraska, United States
\item \Idef{org17}Department of Physics, Aligarh Muslim University, Aligarh, India
\item \Idef{org18}Department of Physics, Pusan National University, Pusan, Republic of Korea
\item \Idef{org19}Department of Physics, Sejong University, Seoul, Republic of Korea
\item \Idef{org20}Department of Physics, University of California, Berkeley, California, United States
\item \Idef{org21}Department of Physics, University of Oslo, Oslo, Norway
\item \Idef{org22}Department of Physics and Technology, University of Bergen, Bergen, Norway
\item \Idef{org23}Dipartimento di Fisica dell'Universit\`{a} 'La Sapienza' and Sezione INFN, Rome, Italy
\item \Idef{org24}Dipartimento di Fisica dell'Universit\`{a} and Sezione INFN, Cagliari, Italy
\item \Idef{org25}Dipartimento di Fisica dell'Universit\`{a} and Sezione INFN, Trieste, Italy
\item \Idef{org26}Dipartimento di Fisica dell'Universit\`{a} and Sezione INFN, Turin, Italy
\item \Idef{org27}Dipartimento di Fisica e Astronomia dell'Universit\`{a} and Sezione INFN, Bologna, Italy
\item \Idef{org28}Dipartimento di Fisica e Astronomia dell'Universit\`{a} and Sezione INFN, Catania, Italy
\item \Idef{org29}Dipartimento di Fisica e Astronomia dell'Universit\`{a} and Sezione INFN, Padova, Italy
\item \Idef{org30}Dipartimento di Fisica `E.R.~Caianiello' dell'Universit\`{a} and Gruppo Collegato INFN, Salerno, Italy
\item \Idef{org31}Dipartimento DISAT del Politecnico and Sezione INFN, Turin, Italy
\item \Idef{org32}Dipartimento di Scienze e Innovazione Tecnologica dell'Universit\`{a} del Piemonte Orientale and INFN Sezione di Torino, Alessandria, Italy
\item \Idef{org33}Dipartimento Interateneo di Fisica `M.~Merlin' and Sezione INFN, Bari, Italy
\item \Idef{org34}European Organization for Nuclear Research (CERN), Geneva, Switzerland
\item \Idef{org35}Faculty of Electrical Engineering, Mechanical Engineering and Naval Architecture, University of Split, Split, Croatia
\item \Idef{org36}Faculty of Engineering and Science, Western Norway University of Applied Sciences, Bergen, Norway
\item \Idef{org37}Faculty of Nuclear Sciences and Physical Engineering, Czech Technical University in Prague, Prague, Czech Republic
\item \Idef{org38}Faculty of Science, P.J.~\v{S}af\'{a}rik University, Ko\v{s}ice, Slovakia
\item \Idef{org39}Frankfurt Institute for Advanced Studies, Johann Wolfgang Goethe-Universit\"{a}t Frankfurt, Frankfurt, Germany
\item \Idef{org40}Gangneung-Wonju National University, Gangneung, Republic of Korea
\item \Idef{org41}Gauhati University, Department of Physics, Guwahati, India
\item \Idef{org42}Helmholtz-Institut f\"{u}r Strahlen- und Kernphysik, Rheinische Friedrich-Wilhelms-Universit\"{a}t Bonn, Bonn, Germany
\item \Idef{org43}Helsinki Institute of Physics (HIP), Helsinki, Finland
\item \Idef{org44}High Energy Physics Group,  Universidad Aut\'{o}noma de Puebla, Puebla, Mexico
\item \Idef{org45}Hiroshima University, Hiroshima, Japan
\item \Idef{org46}Hochschule Worms, Zentrum  f\"{u}r Technologietransfer und Telekommunikation (ZTT), Worms, Germany
\item \Idef{org47}Horia Hulubei National Institute of Physics and Nuclear Engineering, Bucharest, Romania
\item \Idef{org48}Indian Institute of Technology Bombay (IIT), Mumbai, India
\item \Idef{org49}Indian Institute of Technology Indore, Indore, India
\item \Idef{org50}Indonesian Institute of Sciences, Jakarta, Indonesia
\item \Idef{org51}INFN, Laboratori Nazionali di Frascati, Frascati, Italy
\item \Idef{org52}INFN, Sezione di Bari, Bari, Italy
\item \Idef{org53}INFN, Sezione di Bologna, Bologna, Italy
\item \Idef{org54}INFN, Sezione di Cagliari, Cagliari, Italy
\item \Idef{org55}INFN, Sezione di Catania, Catania, Italy
\item \Idef{org56}INFN, Sezione di Padova, Padova, Italy
\item \Idef{org57}INFN, Sezione di Roma, Rome, Italy
\item \Idef{org58}INFN, Sezione di Torino, Turin, Italy
\item \Idef{org59}INFN, Sezione di Trieste, Trieste, Italy
\item \Idef{org60}Inha University, Incheon, Republic of Korea
\item \Idef{org61}Institut de Physique Nucl\'{e}aire d'Orsay (IPNO), Institut National de Physique Nucl\'{e}aire et de Physique des Particules (IN2P3/CNRS), Universit\'{e} de Paris-Sud, Universit\'{e} Paris-Saclay, Orsay, France
\item \Idef{org62}Institute for Nuclear Research, Academy of Sciences, Moscow, Russia
\item \Idef{org63}Institute for Subatomic Physics, Utrecht University/Nikhef, Utrecht, Netherlands
\item \Idef{org64}Institute for Theoretical and Experimental Physics, Moscow, Russia
\item \Idef{org65}Institute of Experimental Physics, Slovak Academy of Sciences, Ko\v{s}ice, Slovakia
\item \Idef{org66}Institute of Physics, Homi Bhabha National Institute, Bhubaneswar, India
\item \Idef{org67}Institute of Physics of the Czech Academy of Sciences, Prague, Czech Republic
\item \Idef{org68}Institute of Space Science (ISS), Bucharest, Romania
\item \Idef{org69}Institut f\"{u}r Kernphysik, Johann Wolfgang Goethe-Universit\"{a}t Frankfurt, Frankfurt, Germany
\item \Idef{org70}Instituto de Ciencias Nucleares, Universidad Nacional Aut\'{o}noma de M\'{e}xico, Mexico City, Mexico
\item \Idef{org71}Instituto de F\'{i}sica, Universidade Federal do Rio Grande do Sul (UFRGS), Porto Alegre, Brazil
\item \Idef{org72}Instituto de F\'{\i}sica, Universidad Nacional Aut\'{o}noma de M\'{e}xico, Mexico City, Mexico
\item \Idef{org73}iThemba LABS, National Research Foundation, Somerset West, South Africa
\item \Idef{org74}Johann-Wolfgang-Goethe Universit\"{a}t Frankfurt Institut f\"{u}r Informatik, Fachbereich Informatik und Mathematik, Frankfurt, Germany
\item \Idef{org75}Joint Institute for Nuclear Research (JINR), Dubna, Russia
\item \Idef{org76}Korea Institute of Science and Technology Information, Daejeon, Republic of Korea
\item \Idef{org77}KTO Karatay University, Konya, Turkey
\item \Idef{org78}Laboratoire de Physique Subatomique et de Cosmologie, Universit\'{e} Grenoble-Alpes, CNRS-IN2P3, Grenoble, France
\item \Idef{org79}Lawrence Berkeley National Laboratory, Berkeley, California, United States
\item \Idef{org80}Lund University Department of Physics, Division of Particle Physics, Lund, Sweden
\item \Idef{org81}Nagasaki Institute of Applied Science, Nagasaki, Japan
\item \Idef{org82}Nara Women{'}s University (NWU), Nara, Japan
\item \Idef{org83}National and Kapodistrian University of Athens, School of Science, Department of Physics , Athens, Greece
\item \Idef{org84}National Centre for Nuclear Research, Warsaw, Poland
\item \Idef{org85}National Institute of Science Education and Research, Homi Bhabha National Institute, Jatni, India
\item \Idef{org86}National Nuclear Research Center, Baku, Azerbaijan
\item \Idef{org87}National Research Centre Kurchatov Institute, Moscow, Russia
\item \Idef{org88}Niels Bohr Institute, University of Copenhagen, Copenhagen, Denmark
\item \Idef{org89}Nikhef, National institute for subatomic physics, Amsterdam, Netherlands
\item \Idef{org90}NRC Kurchatov Institute IHEP, Protvino, Russia
\item \Idef{org91}NRNU Moscow Engineering Physics Institute, Moscow, Russia
\item \Idef{org92}Nuclear Physics Group, STFC Daresbury Laboratory, Daresbury, United Kingdom
\item \Idef{org93}Nuclear Physics Institute of the Czech Academy of Sciences, \v{R}e\v{z} u Prahy, Czech Republic
\item \Idef{org94}Oak Ridge National Laboratory, Oak Ridge, Tennessee, United States
\item \Idef{org95}Ohio State University, Columbus, Ohio, United States
\item \Idef{org96}Petersburg Nuclear Physics Institute, Gatchina, Russia
\item \Idef{org97}Physics department, Faculty of science, University of Zagreb, Zagreb, Croatia
\item \Idef{org98}Physics Department, Panjab University, Chandigarh, India
\item \Idef{org99}Physics Department, University of Jammu, Jammu, India
\item \Idef{org100}Physics Department, University of Rajasthan, Jaipur, India
\item \Idef{org101}Physikalisches Institut, Eberhard-Karls-Universit\"{a}t T\"{u}bingen, T\"{u}bingen, Germany
\item \Idef{org102}Physikalisches Institut, Ruprecht-Karls-Universit\"{a}t Heidelberg, Heidelberg, Germany
\item \Idef{org103}Physik Department, Technische Universit\"{a}t M\"{u}nchen, Munich, Germany
\item \Idef{org104}Politecnico di Bari, Bari, Italy
\item \Idef{org105}Research Division and ExtreMe Matter Institute EMMI, GSI Helmholtzzentrum f\"ur Schwerionenforschung GmbH, Darmstadt, Germany
\item \Idef{org106}Rudjer Bo\v{s}kovi\'{c} Institute, Zagreb, Croatia
\item \Idef{org107}Russian Federal Nuclear Center (VNIIEF), Sarov, Russia
\item \Idef{org108}Saha Institute of Nuclear Physics, Homi Bhabha National Institute, Kolkata, India
\item \Idef{org109}School of Physics and Astronomy, University of Birmingham, Birmingham, United Kingdom
\item \Idef{org110}Secci\'{o}n F\'{\i}sica, Departamento de Ciencias, Pontificia Universidad Cat\'{o}lica del Per\'{u}, Lima, Peru
\item \Idef{org111}Shanghai Institute of Applied Physics, Shanghai, China
\item \Idef{org112}St. Petersburg State University, St. Petersburg, Russia
\item \Idef{org113}Stefan Meyer Institut f\"{u}r Subatomare Physik (SMI), Vienna, Austria
\item \Idef{org114}SUBATECH, IMT Atlantique, Universit\'{e} de Nantes, CNRS-IN2P3, Nantes, France
\item \Idef{org115}Suranaree University of Technology, Nakhon Ratchasima, Thailand
\item \Idef{org116}Technical University of Ko\v{s}ice, Ko\v{s}ice, Slovakia
\item \Idef{org117}Technische Universit\"{a}t M\"{u}nchen, Excellence Cluster 'Universe', Munich, Germany
\item \Idef{org118}The Henryk Niewodniczanski Institute of Nuclear Physics, Polish Academy of Sciences, Cracow, Poland
\item \Idef{org119}The University of Texas at Austin, Austin, Texas, United States
\item \Idef{org120}Universidad Aut\'{o}noma de Sinaloa, Culiac\'{a}n, Mexico
\item \Idef{org121}Universidade de S\~{a}o Paulo (USP), S\~{a}o Paulo, Brazil
\item \Idef{org122}Universidade Estadual de Campinas (UNICAMP), Campinas, Brazil
\item \Idef{org123}Universidade Federal do ABC, Santo Andre, Brazil
\item \Idef{org124}University College of Southeast Norway, Tonsberg, Norway
\item \Idef{org125}University of Cape Town, Cape Town, South Africa
\item \Idef{org126}University of Houston, Houston, Texas, United States
\item \Idef{org127}University of Jyv\"{a}skyl\"{a}, Jyv\"{a}skyl\"{a}, Finland
\item \Idef{org128}University of Liverpool, Liverpool, United Kingdom
\item \Idef{org129}University of Science and Techonology of China, Hefei, China
\item \Idef{org130}University of Tennessee, Knoxville, Tennessee, United States
\item \Idef{org131}University of the Witwatersrand, Johannesburg, South Africa
\item \Idef{org132}University of Tokyo, Tokyo, Japan
\item \Idef{org133}University of Tsukuba, Tsukuba, Japan
\item \Idef{org134}Universit\'{e} Clermont Auvergne, CNRS/IN2P3, LPC, Clermont-Ferrand, France
\item \Idef{org135}Universit\'{e} de Lyon, Universit\'{e} Lyon 1, CNRS/IN2P3, IPN-Lyon, Villeurbanne, Lyon, France
\item \Idef{org136}Universit\'{e} de Strasbourg, CNRS, IPHC UMR 7178, F-67000 Strasbourg, France, Strasbourg, France
\item \Idef{org137}Universit\'{e} Paris-Saclay Centre d'Etudes de Saclay (CEA), IRFU, D\'{e}partment de Physique Nucl\'{e}aire (DPhN), Saclay, France
\item \Idef{org138}Universit\`{a} degli Studi di Foggia, Foggia, Italy
\item \Idef{org139}Universit\`{a} degli Studi di Pavia, Pavia, Italy
\item \Idef{org140}Universit\`{a} di Brescia, Brescia, Italy
\item \Idef{org141}Variable Energy Cyclotron Centre, Homi Bhabha National Institute, Kolkata, India
\item \Idef{org142}Warsaw University of Technology, Warsaw, Poland
\item \Idef{org143}Wayne State University, Detroit, Michigan, United States
\item \Idef{org144}Westf\"{a}lische Wilhelms-Universit\"{a}t M\"{u}nster, Institut f\"{u}r Kernphysik, M\"{u}nster, Germany
\item \Idef{org145}Wigner Research Centre for Physics, Hungarian Academy of Sciences, Budapest, Hungary
\item \Idef{org146}Yale University, New Haven, Connecticut, United States
\item \Idef{org147}Yonsei University, Seoul, Republic of Korea
\end{Authlist}
\endgroup